\documentclass[twocolumn,preprintnumbers,amsmath,amssymb,showpacs,prb,aps]{revtex4}
\usepackage{txfonts}
\usepackage{pifont}
\usepackage{amssymb}
\usepackage{dcolumn}
\usepackage{amsmath}
\usepackage[dvips]{epsfig}
\usepackage {booktabs}

\makeatletter
\def\btt#1{\texttt{\@backslashchar#1}}%
\DeclareRobustCommand\bblash{\btt{\@backslashchar}}%
\makeatother

\begin{document}

\title{Effective minimal model and unconventional spin-singlet pairing in Kagome superconductor  CsV$_{3}$Sb$_{5}$}

\author{Xiao-Cheng Bai$^{1,2}$, Wen-Feng Wu$^{1,2}$}
\thanks{These authors contributed equally. }
\author{Han-Yu Wang$^{1,2}$, Ya-Min Quan$^{1}$, Xian-Long Wang$^{1,2}$}
\author{Zhi Zeng$^{1,2}$}\emph{}
\thanks{zzeng@theory.issp.ac.cn}
\author{Liang-Jian Zou$^{1,2}$}\emph{}
\thanks{zou@theory.issp.ac.cn}

\affiliation{ \it $^1$ Key Laboratory of Materials Physics,Institute of Solid State Physics, HFIPS,
	Chinese Academy of Sciences,Hefei 230031, China\\                
	\it $^2$ Science Island Branch of Graduate School,
	University of Science and Technology of China, Hefei 230026, China\\ 
}
\date{today}  

\begin{abstract}
Recently synthesized Kagome compounds AV$_3$Sb$_5$ attract great attention due to the unusual 
coexistence of the topology, charge density wave and superconductivity. 
 In this {\it Letter}, based on the band structures for CsV$_3$Sb$_5$ in pristine phase, 
 we fit an effective 6-band model for the low-energy processes; utilizing the random phase approximation 
 (RPA) on the effective  minimal model, we obtain the momentum-resolved static spin susceptibility; 
 attributing the spin-fluctuation pairing mechanism, we find that the superconducting pairing strengths 
 increase with the lift of the Coulomb correlation, and the superconductive pairing symmetry is singlet, 
the gap functions are antisymmetric with respect to  the x-axis and the y-axis in the intermediate to 
strong Coulomb correlated regime,  
 indicating the unconventional superconductivity in Kagome compounds AV$_3$Sb$_5$.

\end{abstract}
%\pacs{74.20.Rp, 74.20.Mn, 71.20.−b, 75.40.Gb}
%
\maketitle

%\section{\bf INTRODUCTION} 
%\label{intro}
\noindent{\it Introduction}:
Local spins in a Kagome lattice compound was regarded as most probably forming exotic quantum spin liquid \cite{PhysRevB-83-224413,PhysRevB-103-085128}. Once such Kagome compounds \cite{PhysRevMat-3-094407} become superconductive\cite{PhysRevLett-125-247002,Qiangwei-Yin-37403,PhysRevMaterials-5-034801,PhysRevLett126-247001,PhysRevB-103-224513,Xu-Chen-57402,zhu2021doubledome,nc-12-3645,Shunli-Ni-57403,PhysRevLett-127-207002,PhysRevLett-127-187004,nature-216-599,nature-222-599,nat-mat-CDW,PRB-104-035131,PRB-104-075148,ChaoMu-77402,zhao2021nodal}, the long-term searching relationship between quantum spin liquid phase and unconventional superconductivity seems link with each other;  recently found novel Kagome lattice compounds AV$_3$Sb$_5$  (A = K, Rb, Cs) have been attracting great research interest\cite{PhysRevLett-125-247002,Qiangwei-Yin-37403,PhysRevMaterials-5-034801,PhysRevLett126-247001,PhysRevB-103-224513,Xu-Chen-57402,zhu2021doubledome,nc-12-3645,Shunli-Ni-57403,PhysRevLett-127-207002,PhysRevLett-127-187004,nature-216-599,nature-222-599,nat-mat-CDW,PRB-104-035131,PRB-104-075148,ChaoMu-77402,zhao2021nodal,PhysRevLett-127-046401,PhysRevB-104-195130,wu-2021-magnetic,labollita-2021-tuning,luo-2021-electronic,li-2021-rotation,PhysRevX-11-041010,kang-2021-twofold,cho-2021-emergence}. Through recent great efforts, many physical properties of AV$_3$Sb$_5$, including the electronic structures\cite{PhysRevMat-3-094407,PhysRevLett-127-046401,PhysRevB-104-195130,wu-2021-magnetic,labollita-2021-tuning}, normal-state \cite{PhysRevMat-3-094407,li-2021-rotation} and superconducting-state properties\cite{PhysRevLett-125-247002,Qiangwei-Yin-37403,PhysRevMaterials-5-034801,PhysRevLett126-247001,PhysRevB-103-224513,Xu-Chen-57402,zhu2021doubledome,nc-12-3645,Shunli-Ni-57403,PhysRevLett-127-207002,PhysRevLett-127-187004,nature-216-599,nature-222-599,nat-mat-CDW,PRB-104-035131,PRB-104-075148,ChaoMu-77402,zhao2021nodal}, pressure effect\cite{PhysRevLett126-247001,PhysRevB-103-224513,Xu-Chen-57402,zhu2021doubledome}, magnetic field effect\cite{nc-12-3645,Shunli-Ni-57403,PhysRevLett-127-207002}, STM results \cite{PhysRevLett-127-187004,nature-216-599,nature-222-599,nat-mat-CDW,PRB-104-035131,PRB-104-075148} and ARPES data\cite{PhysRevX-11-041010,kang-2021-twofold,cho-2021-emergence} have gradually become clear. Several theoretical models were also proposed to address the chirality of the charge-density-wave state\cite{Hu-j-p,PRL-127-217601} and superconductivity\cite{PhysRevLett-127-177001}. Nevertheless, a few of essential features remain not revealed, including the effective low-energy model, superconducting pairing force and the pairing symmetry, {\it etc.}. 
\\
 
Due to unique crystal structure of Kagome compounds AV$_3$Sb$_5$, the electronic structures of normal-state AV$_3$Sb$_5$ are rather complicated. The band structures of the pristine phase of CsV$_3$Sb$_5$\cite{PhysRevLett-127-046401}, calculated by the density functional theory (DFT), predicted that seven bands cross the Fermi energy E$_{F}$ and are composed of V 3$d$ orbitals and Sb 5$p$ orbitals, which shows typical multi-orbital character. Since the unit cell of Kagome lattice contains three V sites in the pristine phase, at the same time, the metallic ground state in AV$_3$Sb$_5$ and partial filled V 3$d$ orbitals also imply the multi-orbital character in the low-energy model. These inevitably bring difficulty in constructing an effective multiorbital model for superconducting AV$_3$Sb$_5$. At present the microscopic origin of the charge density wave is generally attributed to the {\it van Hove} singularity, whereas, to understand the essential superconductive nature, a proper effective model correctly describing the low-energy processes plays key roles and is highly expected.
\\

So-far experimental data strongly support unconventional  superconductive pairing mechanism in AV$_3$Sb$_5$: conventional e-ph coupling mechanism is not enough to account for the superconductive microscopic origin, neither the transition temperatures estimated by the McMillan's formula for RbV$_3$Sb$_5$ and CsV$_3$Sb$_5$ considerably deviate from the experimental T$_C$ \cite{PhysRevLett-127-046401}, nor the experimental T$_C$ violates the expectation of the conventional e-ph mechanism when A varies from K, Rb to Cs. AV$_3$Sb$_5$ superconductors also display many features of unconventional superconductivity: with the increase of hydrostatic pressure, the P-T phase diagram demonstrates two superconductive regions \cite{PhysRevLett126-247001,zhu2021doubledome}, similar to the two-dome superconductive phase diagrams in iron-based superconductors\cite{iron-twodome-nature-67-483,PhysRevB-97-020508}. Though the local magnetic moment was argued to be absent \cite{Kenney-2021}, the observation of the anomalous Hall effect in CsV$_3$Sb$_5$  \cite{sciadv-abb6003} strongly favors of the presence of local magnetic field, hence the local spins. The first-principles electronic structure calculations also suggested there exists local magnetic moments in AV$_3$Sb$_5$ \cite{PhysRevB-104-195130, wu-2021-magnetic},  implying that the spin-orbital fluctuations mechanism is the possible pairing origin, since not only the system posses local magnetic moments and multiorbital character, but also Kagome lattice structure favors strong spin frustrations and fluctuations. These eager further deep theoretical investigations.
\\

On the other hand, the details of the superconductive nature in AV$_3$Sb$_5$ are not clear. A very recent study\cite{PhysRevLett-127-177001} suggested that the superconducting electrons in CsV$_3$Sb$_5$ are f-wave triplet pairing: {\it i.e.} the superconductive Cooper pairs are spatial f-wave, and spin parallel or triplet. Such a prediction was inconsistent with the temperature-dependence Knight shift experiment\cite{ChaoMu-77402}. One also finds that in similar sixfold-symmetric superconductor Na$_x$CoO$_2$.yH$_2$O, an earlier theoretical investigation suggested to be f-wave pairing symmetry \cite{naturephysics-91-1}, however, the experiments supported that the pairing symmetry of  superconducting cobaltates is singlet d-wave \cite{MATANO-2007-687}. This arises the puzzle whether the Cooper pairs with high angular momentum could stably exist in AV$_3$Sb$_5$.  In this Letter, based on our calculated band structures for CsV$_3$Sb$_5$ in the pristine phase, we fit an effective low-energy 6-band model; utilizing the random phase approximation (RPA) on the effective minimal model, we show that the superconducting pairing strengths increase with the increase of Coulomb correlation, and the superconductive pairing symmetry is singlet d$_{xy}$-wave-like. The present theoretical results favor our understand on the unconventional superconductivity in Kagome compounds AV$_3$Sb$_5$.
\\

 \begin{figure}[htp]
	\centering
\includegraphics[angle=0, width=0.49 \columnwidth]{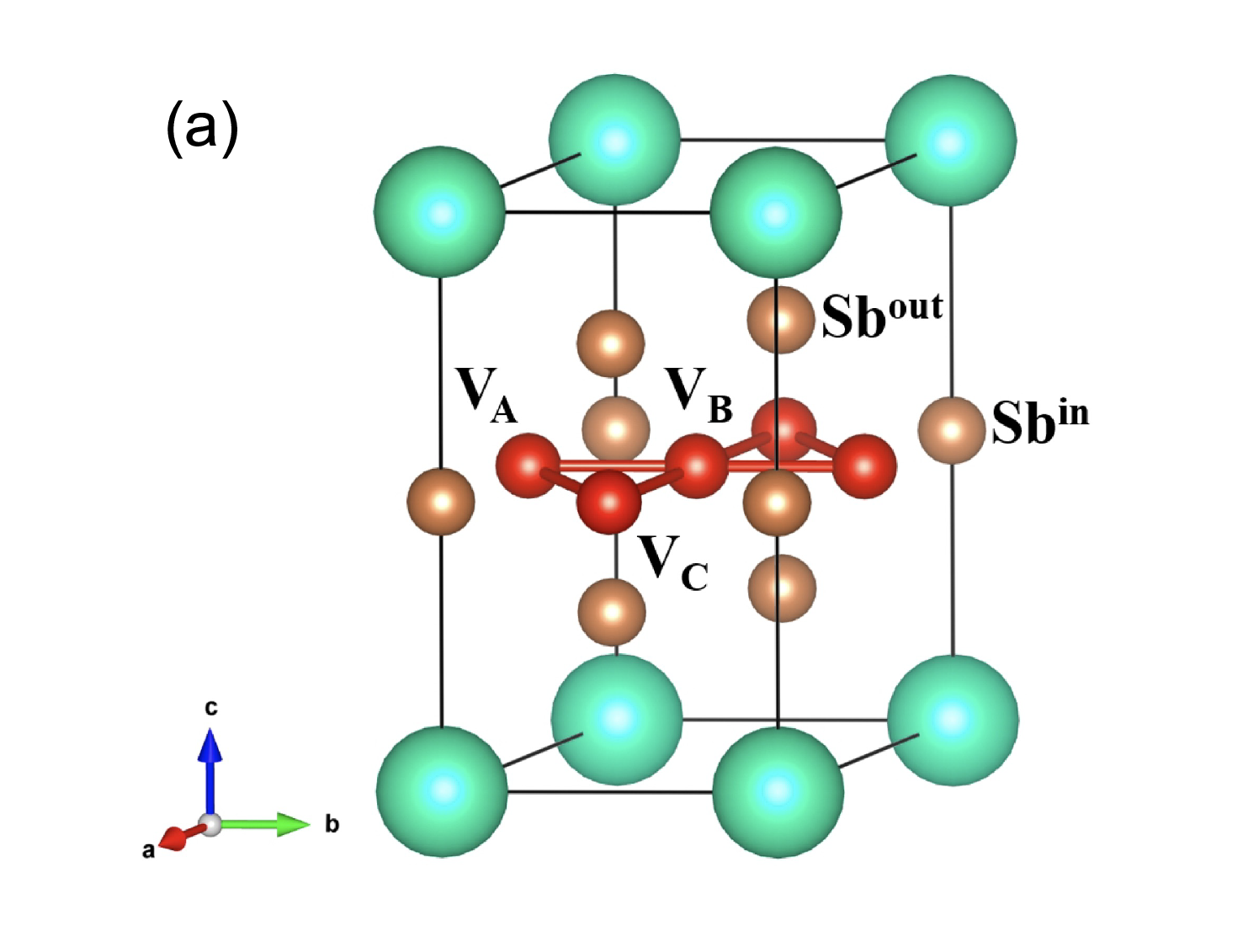}
\includegraphics[angle=0, width=0.49 \columnwidth]{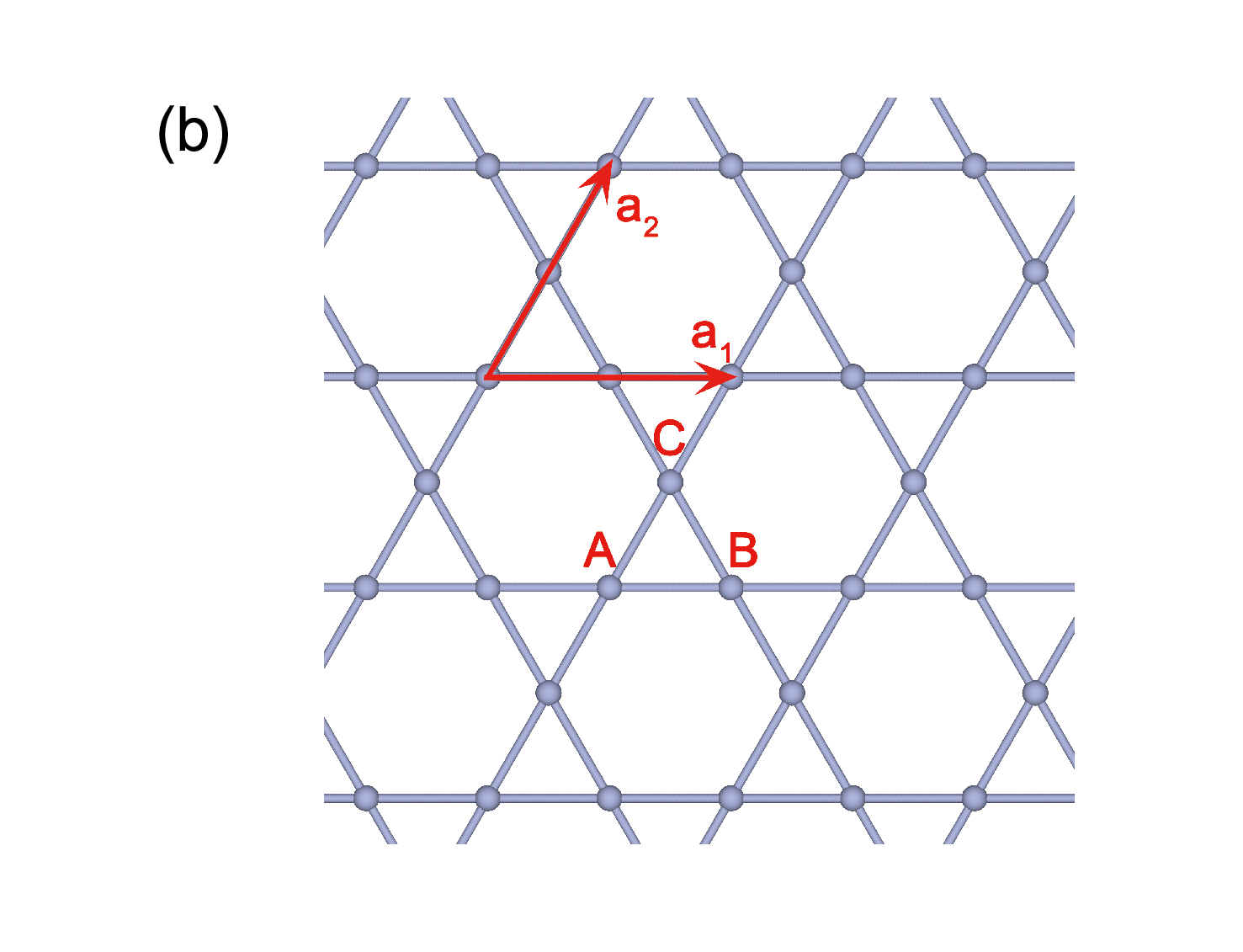}

	\caption{(a) The crystal sttructure of pristine phase CsV$_{3}$Sb$_{5}$. Red, yellow and green balls represent V, Sb and Cs. Sb$^{\rm{in}}$ and Sb$^{\rm{out}}$ indicate that Sb is in the same plane as V and out of the V-Sb plane, respectively. (b)The basic vectors and different V sites on the Kagome plane.} 	   
	\label{Fig.1}
\end{figure}

 \begin{figure}[ht]
	\centering
\includegraphics[angle=0, width=0.50\columnwidth]{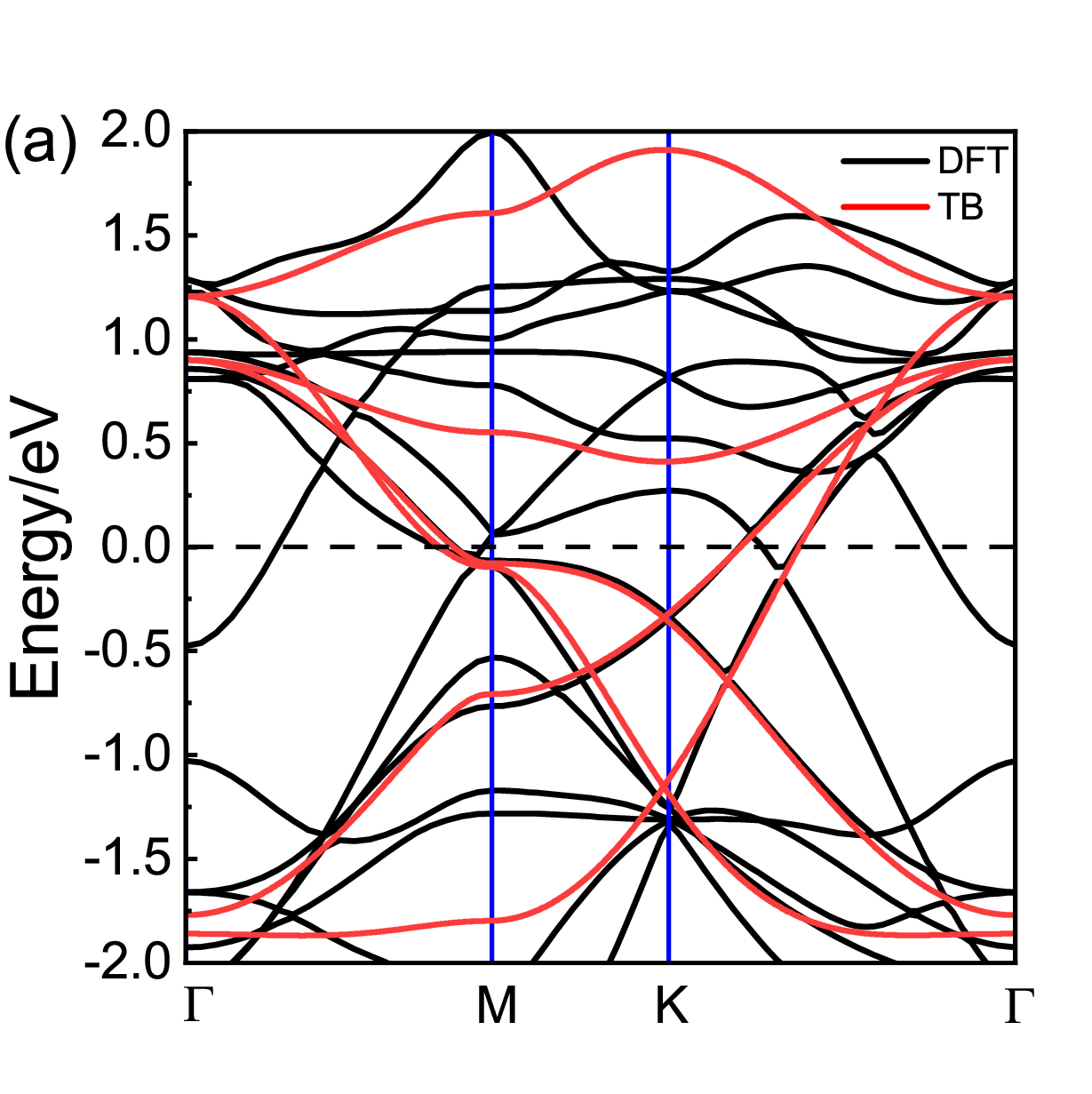}
\includegraphics[angle=0, width=0.47 \columnwidth]{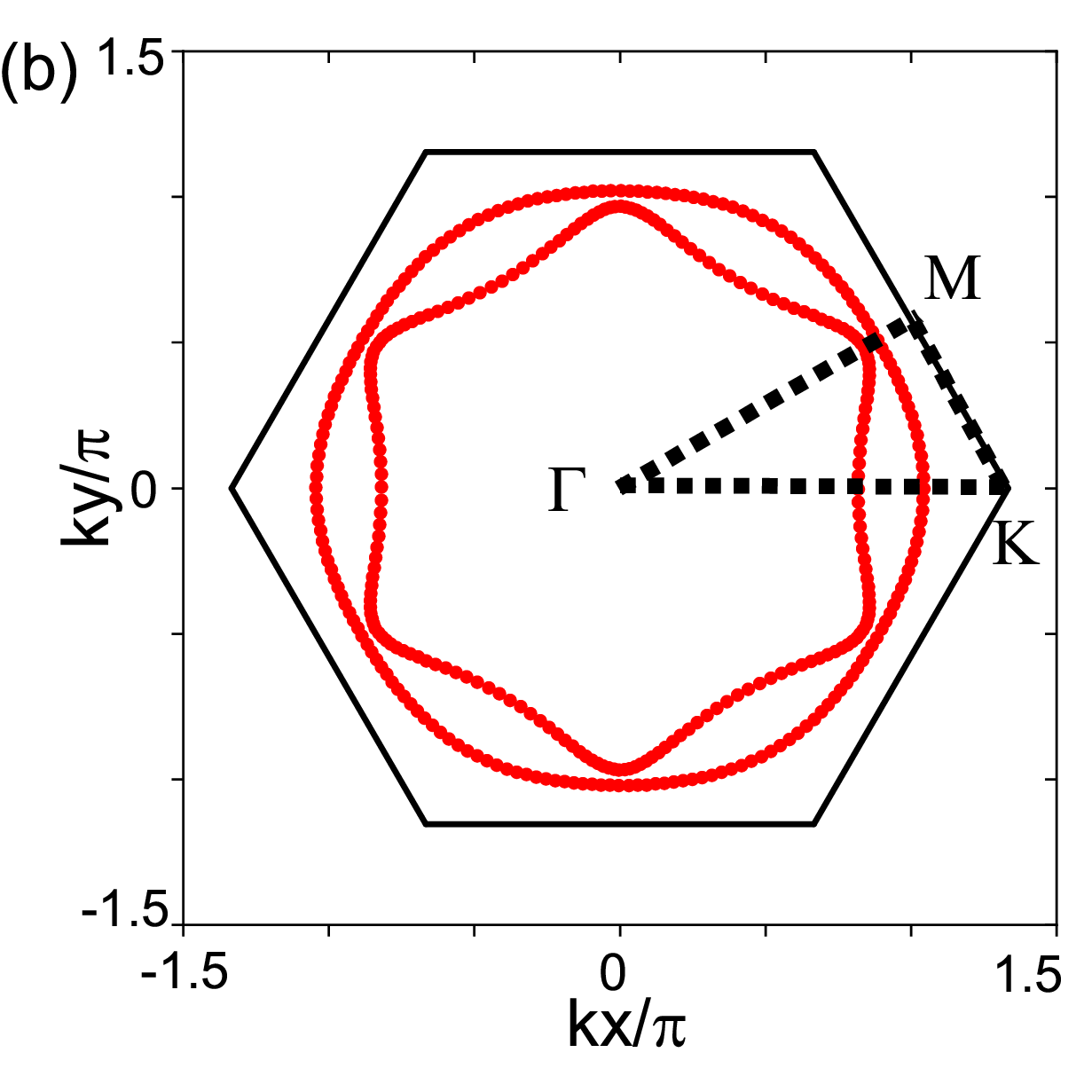}

	\caption{(a) The band structures of pristine CsV$_{3}$Sb$_{5}$ (black lines) and fitting by six-band tight-binding model(red lines); (b) Fermi surface of the six band tight-binding model.} 	   
	\label{Fig.2}
\end{figure}

%CCCCCCCCCCCCCCCCCCCCCCCCCCCCCCCCCCCCCCCCCCCCCCCCCCCCCCCCCCCCCCC

%\section {\bf Effective minimal model Hamiltonian}
%\label{model1}

\noindent{\it Effective minimal model Hamiltonian}:
To investigate the many body effects and superconductivity in AV$_{3}$Sb$_{5}$, one should first construct an effective low-energy model. To this end, we start to perform the first-principles electronic structures calculations to obtain the band structures of CsV$_{3}$Sb$_{5}$. The details of the method and the calculations are given in the {\it Supplementary Materials}. Fig. \ref{Fig.1}(a) shows the crystal structure of the layered CsV$_{3}$Sb$_{5}$. Recent resistivity measurement on CsV$_{3}$Sb$_{5}$ \cite{PhysRevLett-125-247002} showed remarkable anisotropy of the $ab$-plane to the $c$-axis, indicating that  the interlayer V-V hybridization is rather small. Throughout this paper we confine our study on the V-Sb Kagome plane, as seen in Fig. \ref{Fig.1}(b).
\\
 
Our band structures and Fermi surfaces are in agreement with available literature \cite{labollita-2021-tuning,luo-2021-electronic}.
There are three distinct Fermi surface sheets at k$_z$=0 :
(i) a central ring sheet around $\Gamma$ contributed by Sb-$p_{z}$ orbitals, (ii) a hexagonal sheet composed of V-$d_{xy}$, V-$d_{x^{2}-y^{2}}$, and V-$d_{z^{2}}$, and (iii) two additional sheets composed of V-$d_{xz}/d_{yz}$ orbitals\cite{labollita-2021-tuning,luo-2021-electronic}.  
In fitting the band structures with an effective low-energy tight-binding model, 
we consider the large hexagonal Fermi pocket (ii) and one of the Fermi surface sheets (iii), to which the corresponding energy bands are hybridized weakly with the $p$ orbitals of Sb\cite{labollita-2021-tuning}, and then construct a six-band  tight-binding (TB) model with two local orbitals on each V site.
By constructing localized Wannier functions, the hopping integrals between different V orbitals $\alpha$ and $\beta$ ($\alpha=xz,yz$, $\beta=xy,3z^{2}-r^{2},x^{2}-y^{2}$) are zeros, see the TABLE S1 in {\it Supplementary Materials}, indicating that interorbital hybridization between the $\alpha$ and $\beta$ orbits of vanadiums is vanishing small, so we neglect the interorbital hopping terms in the tight-binding model.
The fitted tight-binding bands and the full electronic structures of CsV$_{3}$Sb$_{5}$ are the red and black lines in Fig. \ref{Fig.2}(a), respectively. 
\\

The fitted six pack-band tight-binding model for CsV$_{3}$Sb$_{5}$ can be written as
\begin{eqnarray}
\label{eq:tight-binding model}
H_{TB}=\sum_{k,i,j,\alpha,\sigma}  \xi_{ij,\alpha}(k)c^{\dagger}_{ki \alpha \sigma}c_{kj \alpha \sigma} 
\end{eqnarray}
where the inequivalent V sites $i,j=A,B,C$, the orbital indices $\alpha,\beta=1,2$, and the hopping matrix elements $\xi_{ij,\alpha}(k)=\left [(\varepsilon_{\alpha}+t^{\prime\prime}_{\alpha}\phi^{3}(k)-\mu)\delta_{ij}+t_{\alpha}\phi^{1}_{ij}(k)+t^{\prime}_{\alpha}\phi^{2}_{ij}(k) \right]$, 
where $\phi^{n}$ denotes the lattice structure factor defined in {\it Supplementary Materials}.
 The $\varepsilon_{\alpha}$ denotes the energy level of the orbital $\alpha$ and $\mu$ is the chemical potential. The operator $c^{\dagger}_{k,i,\alpha\sigma}$ ($c_{k,i,\alpha\sigma}$) creates (annihilates) an electron with momentum $k$ and spin $\sigma$ of sublattice $i$ in orbital $\alpha$.
 The $t_{\alpha}$, $t^{\prime}_{\alpha}$ and $t^{\prime\prime}_{\alpha}$ denote intra-orbital hopping integrals for the first, second and third nearest neighbors respectively. The hopping parameters are given in {\it Supplementary Materials}, see Table S3. Within the present six-band  model, the Fermi surface is plotted in Fig. \ref{Fig.2}(b). The corresponding electron filling number is $n=5.45$ to match the chemical potential of  CsV$_{3}$Sb$_{5}$ .
\\

To investigate the spin fluctuations and superconducting pairing properties in CsV$_{3}$Sb$_{5}$, we consider the on-site Coulomb interaction as following
\begin{eqnarray}
\label{eq:Hamiltonian1}
H_{int}&=&U\sum_{li,\alpha}n_{li\alpha\uparrow}n_{li\alpha\downarrow}
   +\sum_{li,\sigma,\sigma^{\prime},\alpha >\beta}
   U^{\prime} n_{li\alpha\sigma}n_{li\beta\sigma^{\prime}}        \nonumber\\
   &-&J_{H}\sum_{li,\alpha >\beta} c^{\dagger}_{li\alpha\uparrow}c_{li\alpha
   \downarrow}c^{\dagger}_{li\beta\downarrow}c_{li\beta\uparrow}
   +J_{P}\sum_{li,\alpha \neq\beta} c^{\dagger}_{li\alpha\uparrow}
   c^{\dagger}_{li\alpha\downarrow}c_{i\beta
   \downarrow}c_{li\beta\uparrow}, \nonumber\\
\end{eqnarray}
here $l$ denote the $l-$th unit cell. The on-site intra- and interorbital Coulomb repulsions 
are denoted by $ U $ and $ U^{\prime} $. $J_{H}$ and $ J_{P} $ are the Hund's rule exchange and pair-hopping term, respectively.
Throughout this paper, we set the Couloub and Hund's interaction parameters $ U^{\prime}=U-2J_{H} $ and $ J_{H}=J_{P}=U/8 $, which satisfy spin rotational invariance. The fact that the DFT band structures are in good agreement with the angle-resolved photoemission spectroscopy (ARPES) data\cite{PhysRevLett-125-247002}, demonstrates that the electronic correlation in CsV$_{3}$Sb$_{5}$ is not very strong. 
\\     

Within the random phase approximation (RPA), the singlet (s) and triplet (t) pairing vertices arising from spin and charge fluctuations for the multiorbital case \cite{Takimoto_2002,Takimoto-2003,Kubo-PhysRevB-75-224509-2007} are given by
 \begin{eqnarray}
\label{eq:scattering vertex}
\Gamma^{pq,s}_{st}(k,k^{\prime},\omega)&=&[ \frac{3}{2}U^{s}\chi^{RPA}_{S}(k-k^{\prime},\omega)U^{s} \nonumber \\ 
&-&\frac{1}{2}U^{c}\chi^{RPA}_{O}(k-k^{\prime},\omega)U^{c}+\frac{1}{2}(U^{s}+U^{c})]^{qt}_{ps}, \nonumber\\
\\
\Gamma^{pq,t}_{st}(k,k^{\prime},\omega)&=&[ -\frac{1}{2}U^{s}\chi^{RPA}_{S}(k-k^{\prime},\omega)U^{s} \nonumber \\ 
&-&\frac{1}{2}U^{c}\chi^{RPA}_{O}(k-k^{\prime},\omega)U^{c}+\frac{1}{2}(U^{s}+U^{c})]^{qt}_{ps}, \nonumber\\
\end{eqnarray} 
here $U^{s}$ and $U^{c}$ represent the $12\times 12$ matrices in the V orbital space, and $\chi^{RPA}_{S}$ and $\chi^{RPA}_{O}$ are the ones RPA spin  and orbital (charge)  susceptibility matrix given in the Sec. III of {\it Supplementary Materials}.
\\

By projecting the zero frequency $(\omega=0)$ pairing vertex into the band space \cite{Graser_2009,Scalapino-RevModPhys-84-1383}, we obtain
 \begin{eqnarray}
\label{eq:pairing vertex}
\Gamma^{s/t}_{i,j}(k,k^{\prime})&=&\sum_{s,t,p,q} a^{s,*}_{i}(k) a^{t,*}_{i}(-k) {\rm Re} \left[ \Gamma^{pq,s/t}_{st}(k,k^{\prime},0) \right] \nonumber\\
&& \times a^{p}_{j}(k^{\prime}) a^{q}_{j}(-k^{\prime}).
\end{eqnarray} 
The $\Gamma^{s}_{i,j}(k,k^{\prime})$ (or $\Gamma^{t}_{i,j}(k,k^{\prime})$) determines the scatterings of two electrons of opposite (or same) spin from the state $(k,-k)$ on the Fermi surface sheet $C_{i}$ to the state $(k^{\prime},-k^{\prime})$ on the Fermi surface sheet $C_{j}$. By means of a mean-field decouping of the interaction, the gap equation in singlet and triplet channels can be obtained and expressed as
 \begin{eqnarray}
\label{eq:gap equation}
\Delta^{i,s/t}_{k}&=& -\frac{1}{2N}\sum_{j,k^{\prime}}V^{s/t}_{i,j}(k,k^{\prime}) \frac{\Delta^{j,s/t}_{k^{\prime}}}{\Omega^{j,s/t}_{k^{\prime}}}\tanh(\frac{\beta\Omega^{j,s/t}_{k^{\prime}}}{2})
\end{eqnarray} 
with
 \begin{eqnarray}
\label{eq:bogoliubov quasi-particle energy}
\Omega^{i,s/t}_{k}&=& \sqrt{[E_{i}(k)]^{2}+|\Delta^{i,s/t}_{k}|^{2}}, \\
V^{s}_{i,j}(k,k^{\prime})&=&\frac{1}{2}[\Gamma^{s}_{i,j}(k,k^{\prime})+\Gamma^{s}_{i,j}(-k,k^{\prime})],\\
V^{t}_{i,j}(k,k^{\prime})&=&\frac{1}{2}[\Gamma^{t}_{i,j}(k,k^{\prime})-\Gamma^{t}_{i,j}(-k,k^{\prime})].
\end{eqnarray} 
It is worthy noting that $\Delta^{i,s}_{k}$ is an even function with respect to {\bf k} and $\Delta^{i,t}_{k}$ an odd function. The superconducting instability appears at temperature. When the temperature just below T$_{c}$ the gap function $\Delta^{i,s/t}_{k}$ are expected to be very small, and hence we study the superconducting instability by solving the linearized gap equations \cite{Graser_2009,Scalapino_2019-PhysRevB-99-224515}
\begin{eqnarray}
\label{eq:linearized gap equation}
-\frac{1}{V_{G}}\sum_{j} \oint \frac{dk^{\prime}}{|v_{F_{j}}(k^{\prime})|}V^{s/t}_{i,j}\left(k,k^{\prime}\right) 
g_{j}^{\alpha}\left(k^{\prime}\right)=\lambda_{\alpha} g_{i}^{\alpha}\left(k\right).
\end{eqnarray}
Here the eigenvalues $\lambda_{\alpha}$ are the superconducting pairing strengths and $g_{i}^{\alpha}(k)$ the corresponding eigenfunctions. $V_{G}$ is the area of a Brillouin zone, $i$ and $j$ are the band indices of Fermi surface vectors $k$, $k'$, respectively, $|v_{F_{j}}( k^{\prime} )|$ is the magnitude of the Fermi velocity. The largest pairing strength $\lambda_{\alpha}$ determines the highest transition temperature and the corresponding eigenfunction $g_{i}^{\alpha}(k)$ shows the symmetry of the gap function. Throughout this paper we perform the calculations at temperature of $k_{B}T=0.002$ eV.
\\

\begin{figure}[htpb]
	\centering
\includegraphics[angle=0, width=0.80 \columnwidth]{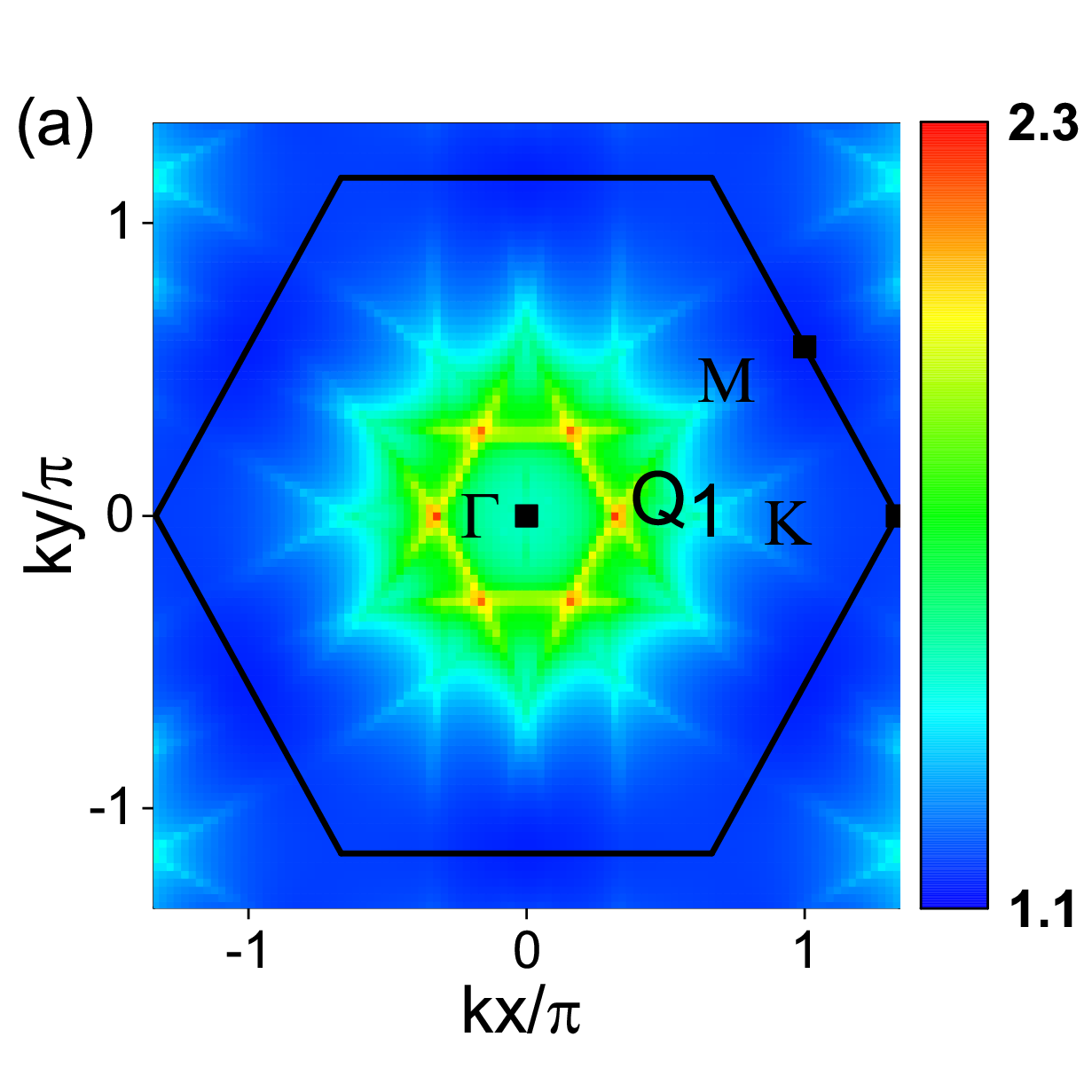}
\includegraphics[angle=0, width=0.80 \columnwidth]{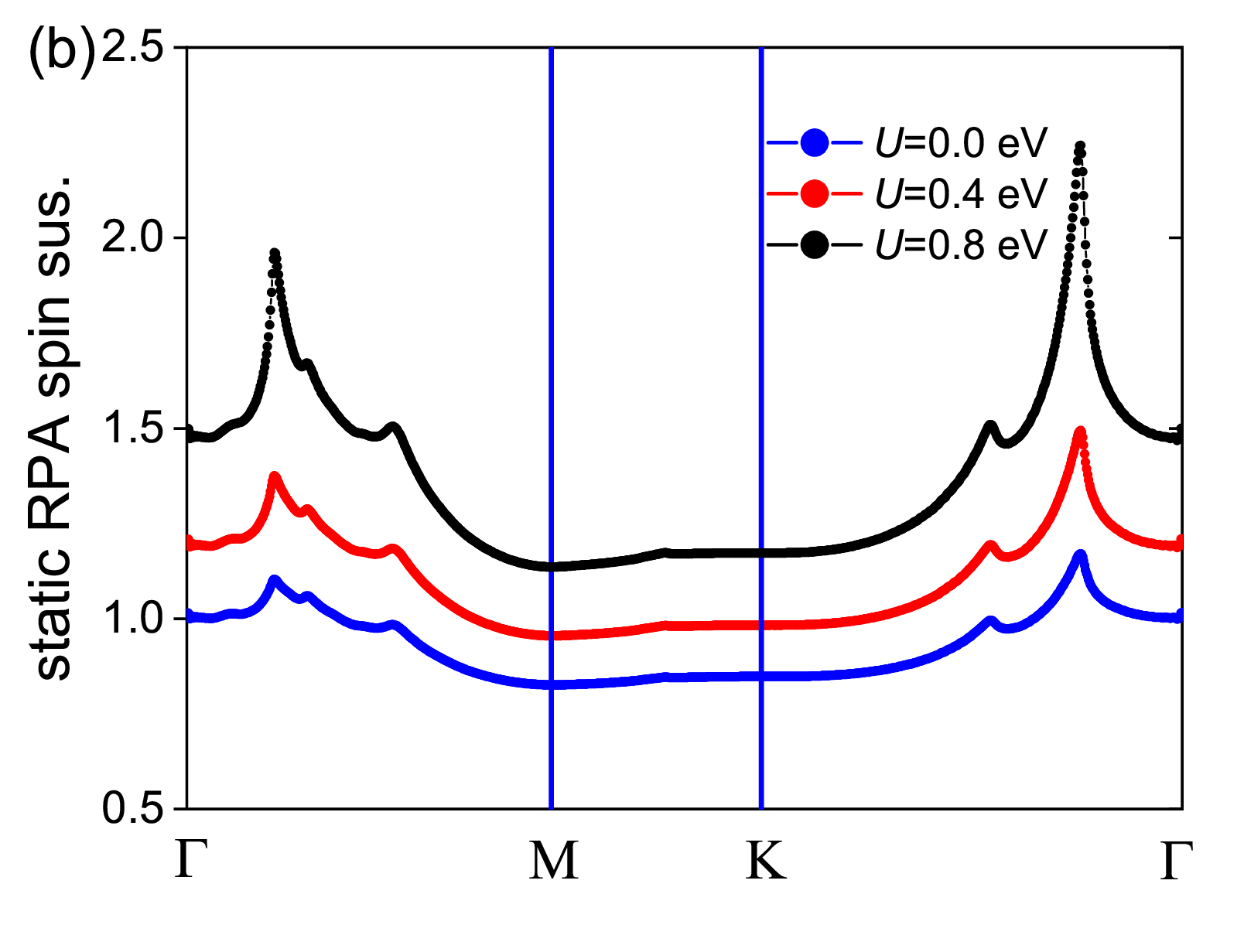}
	\caption{(Color online) (a) Momentum-resolved spin susceptibility of pristine CsV$_{3}$Sb$_{5}$ in the first Brillouin zone for $U=0.8$ eV, and (b) the RPA spin susceptibility along the path of high-symmetry points for different $U$.} 	   
	\label{Fig.sus}
\end{figure}

\noindent{\it Momentum-resolved Spin Susceptibility}:
The distribution of spin susceptibility in momentum space may not only disclose the spin fluctuation modes or the magnetic order, but also provide the superconductive pairing information. The RPA spin susceptibility of the effective minimal model for pristine CsV$_{3}$Sb$_{5}$ shows six peaks near ${\bf Q}_{1}  \approx \frac{1}{4}\Gamma K$ and other five sixfold-symmetric wavevectors, as seen the red points in Fig. \ref{Fig.sus}(a).  In the lines connecting these maxima, the magnitude of the spin susceptibility is also very large, as seen the yellow lines in Fig. \ref{Fig.sus}(a). These indicate that there exist multi-mode antiferromagnetic or spin-density-wave fluctuations. We also find that these peaks are enhanced with the increase of the Coulomb correlation U, as seen in Fig. \ref{Fig.sus}(b).  
% As seen the Fermi surfaces in Fig.2(b), a  fractional of Fermi surface is connected with the counter part after the wavevector shifts of $\bg 
% q_{1}$ and $\bf q$_{2}, showing that the Fermi surface nesting occurs in the electronic structures.
% The nesting of the electronic structures contributes these peaks in the static spin susceptibility of the pristine CsV$_{3}$Sb$_{5}$???.
%
\\

Note that the e-ph coupling mechanism could not account for the origin of the superconductivity in CsV$_{3}$Sb$_{5}$, the spin-fluctuation
mechanism seems a plausible candidate. In the theory of spin-fluctuation mediated superconductivity,
the singlet pair scattering $\Gamma^{s}_{i,j}(k,k^{\prime})$ of a Cooper pair at k to k$^{\prime}$ is proportional to the RPA spin susceptibility\cite{Hirschfeld-PhysRevB-95-174504,Scalapino-RevModPhys-84-1383}, one expects that the gap function on different parts Fermi surface connected by ${\bf q}_{1} \approx \frac{1}{4}\Gamma K$ 
and other sixfold-symmetric wavevectors has opposite signs, according to the linearized gap equations.
\\

%\section{Superconductive pairing strengths}

\label{results}
 \begin{figure}[htpb]
	\centering
\includegraphics[angle=0, width=0.89 \columnwidth]{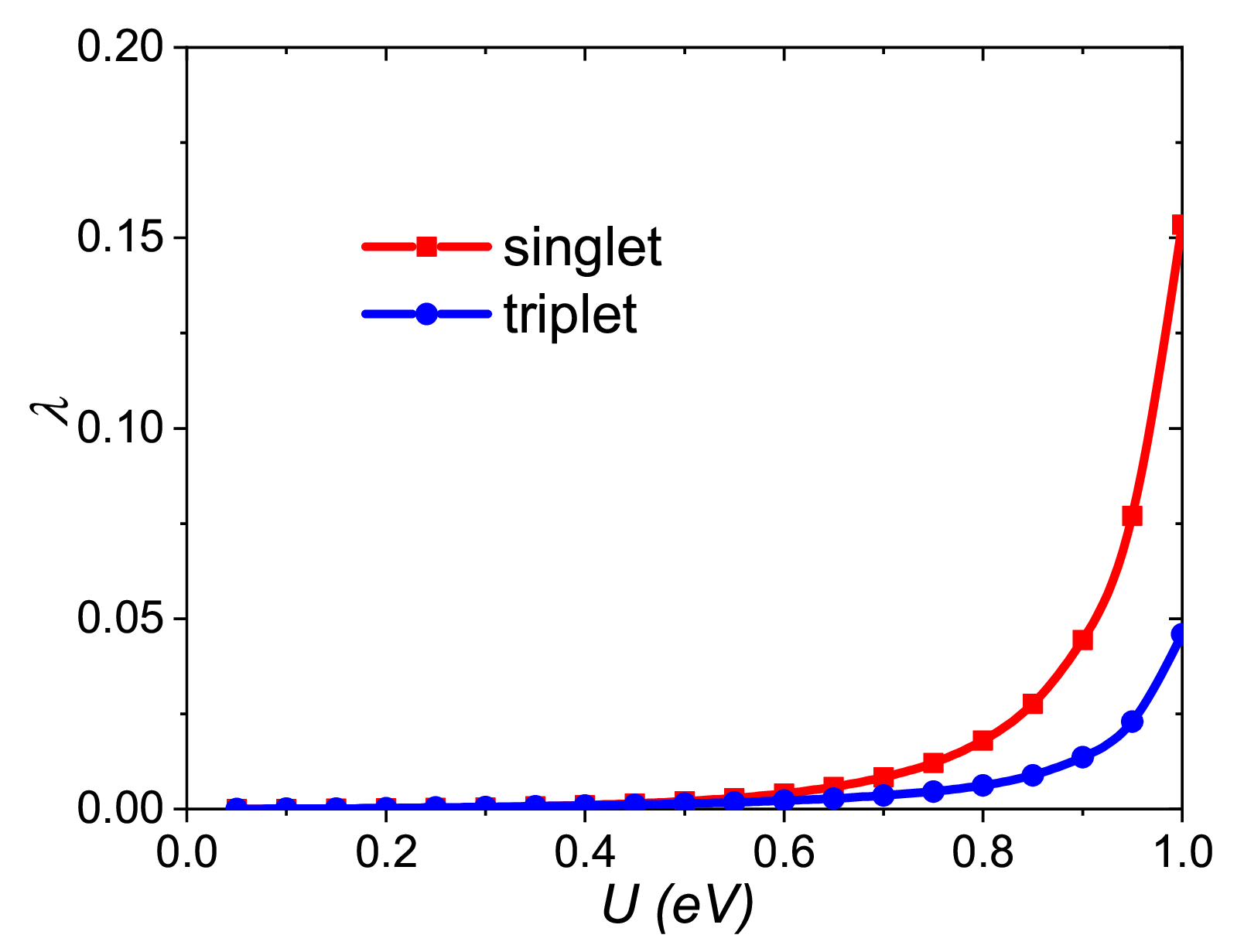}
	\caption{(Color online) The  pairing strength versus interaction $U$, red line for singlet channel and blue line for triplet channel, show that singlet pairing dominate the phase diagram in the range of $U>0.4$ eV.} 	   
	\label{Fig.pairing}
\end{figure}
 \begin{figure}[htpb]
	\centering
\includegraphics[angle=0, width=0.49 \columnwidth]{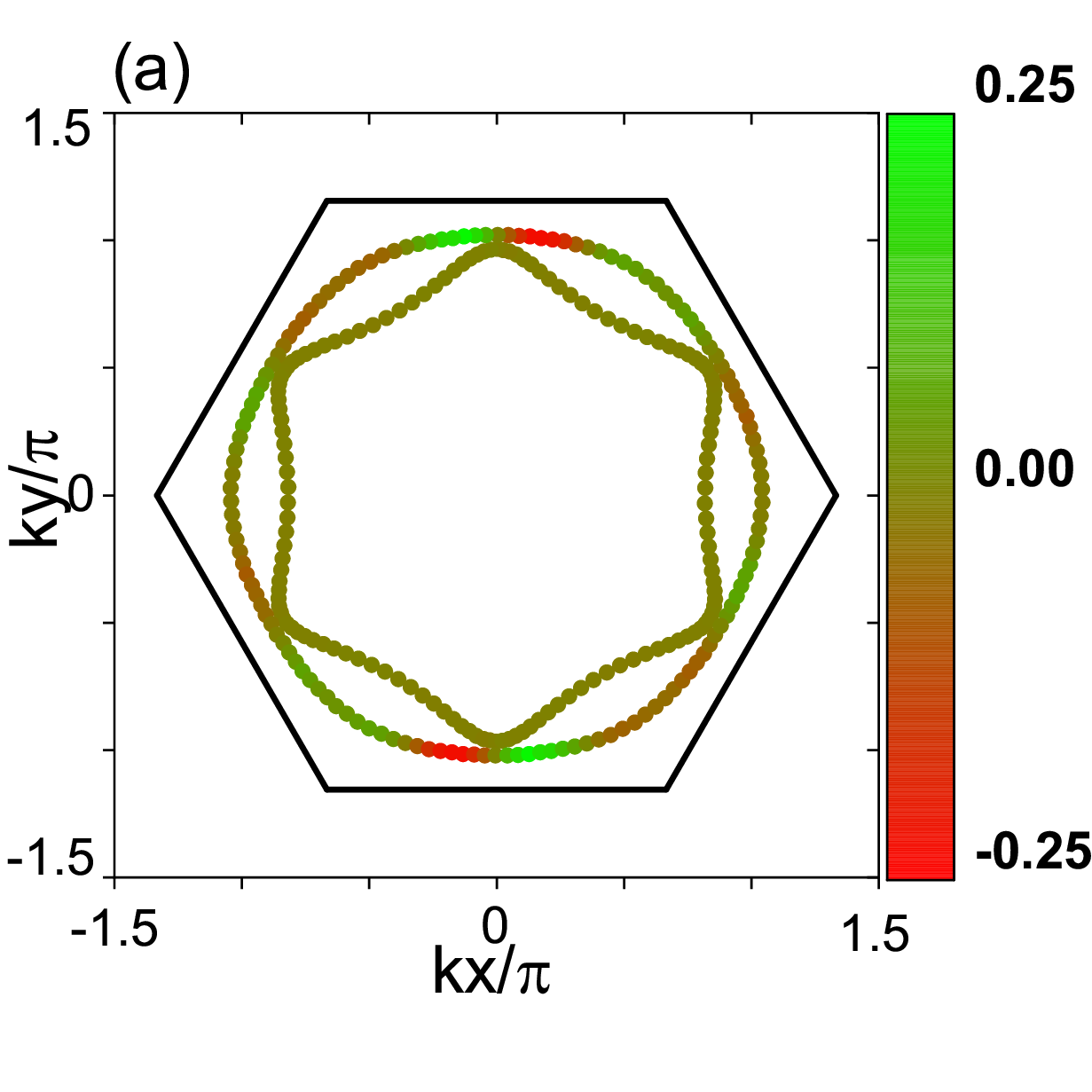}
\includegraphics[angle=0, width=0.49 \columnwidth]{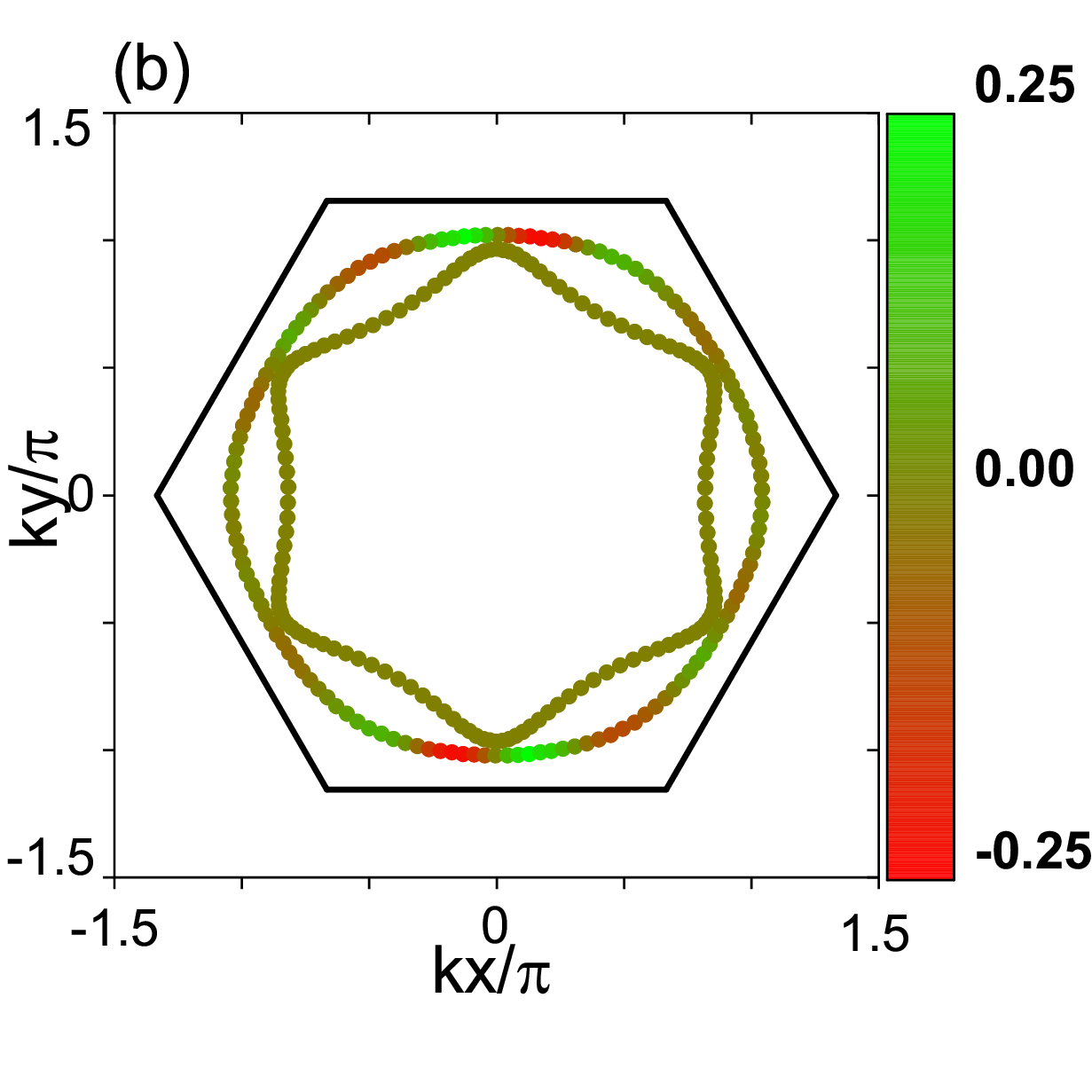}
	\caption{The dominant singlet pairing gap function at $U=0.5$ eV (a) and 0.8 eV (b), respectively.} 	   
	\label{Fig.gap function}
\end{figure}

 \begin{figure}[htpb]
	\centering
\includegraphics[angle=0, width=0.49 \columnwidth]{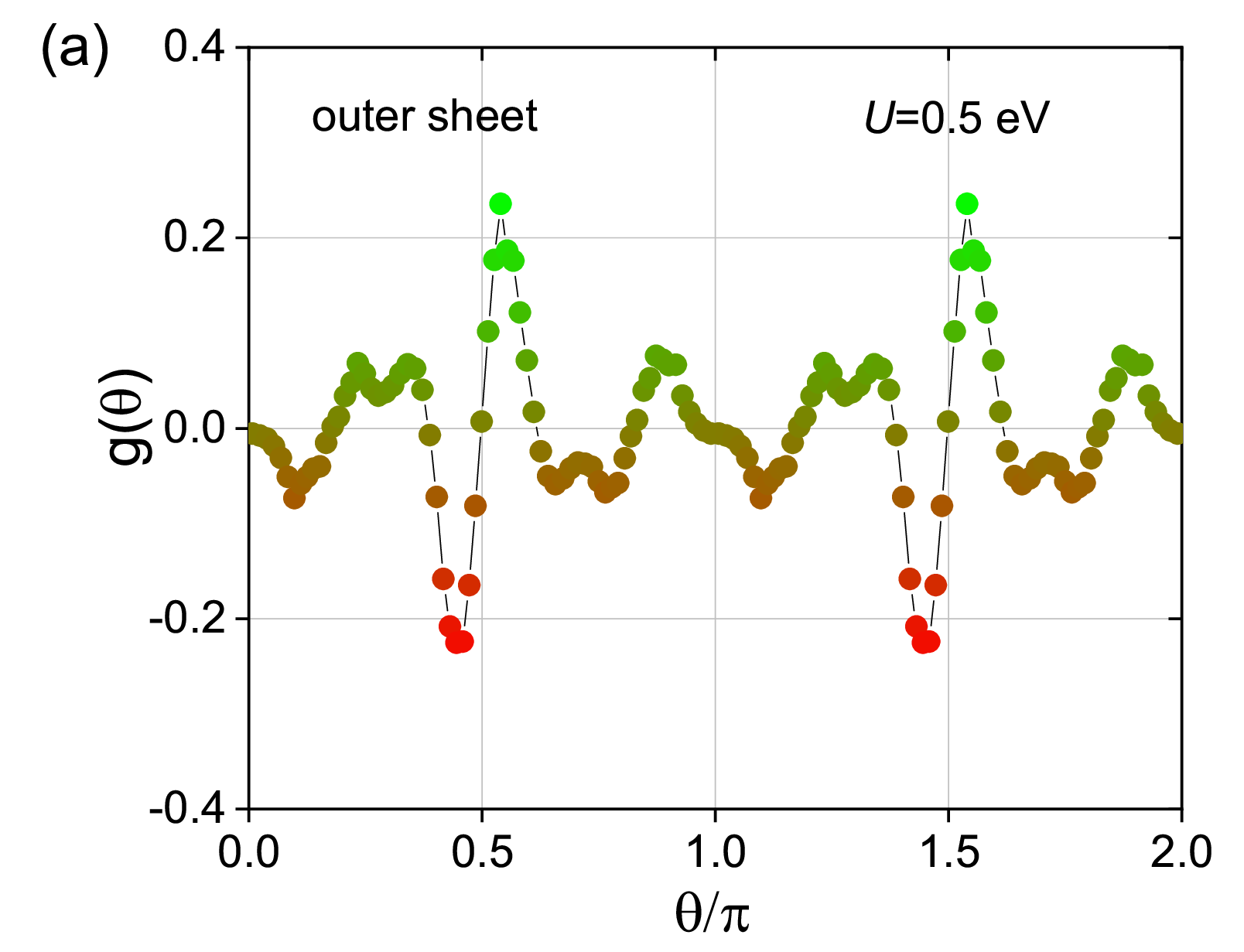}
\includegraphics[angle=0, width=0.49 \columnwidth]{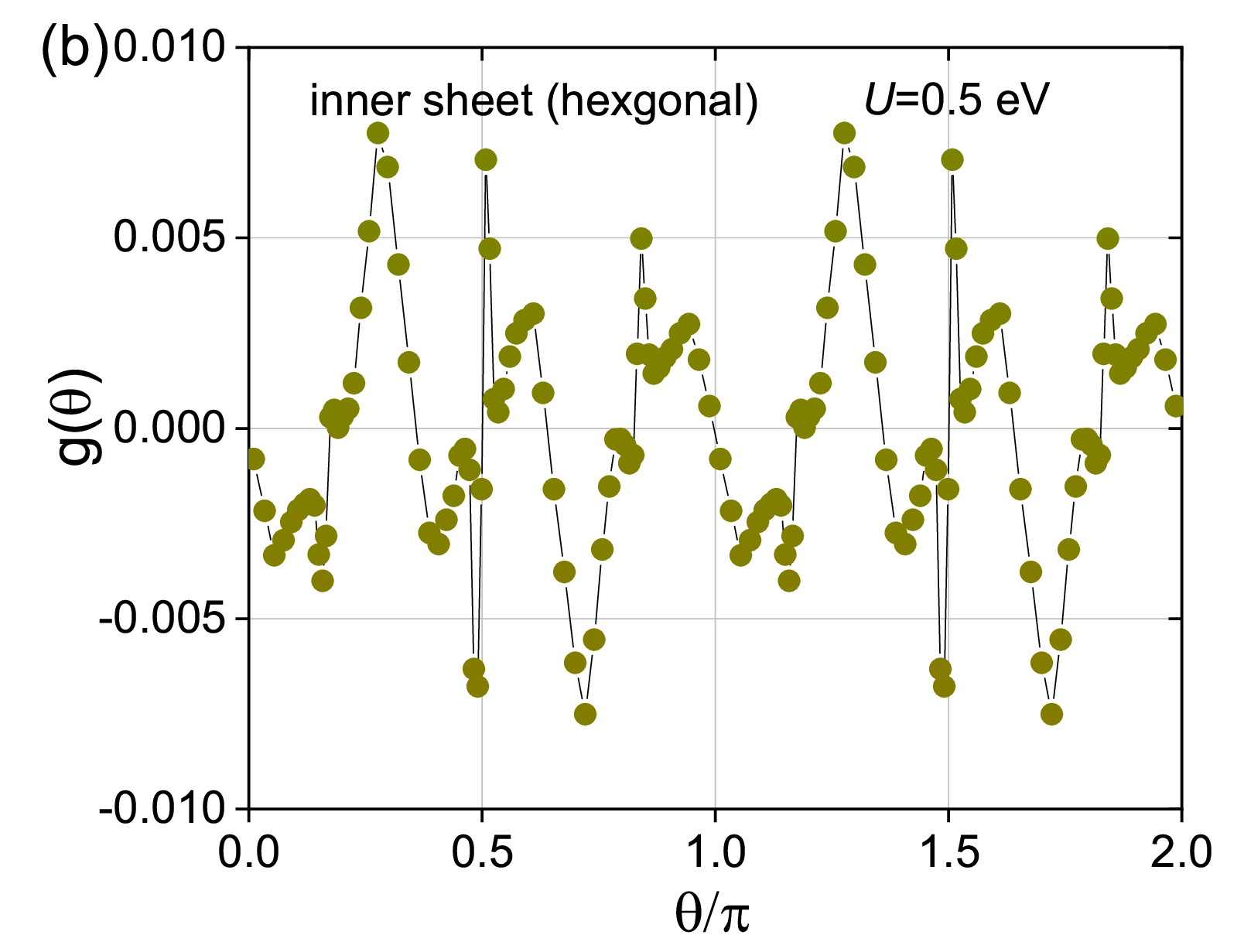}
\includegraphics[angle=0, width=0.49 \columnwidth]{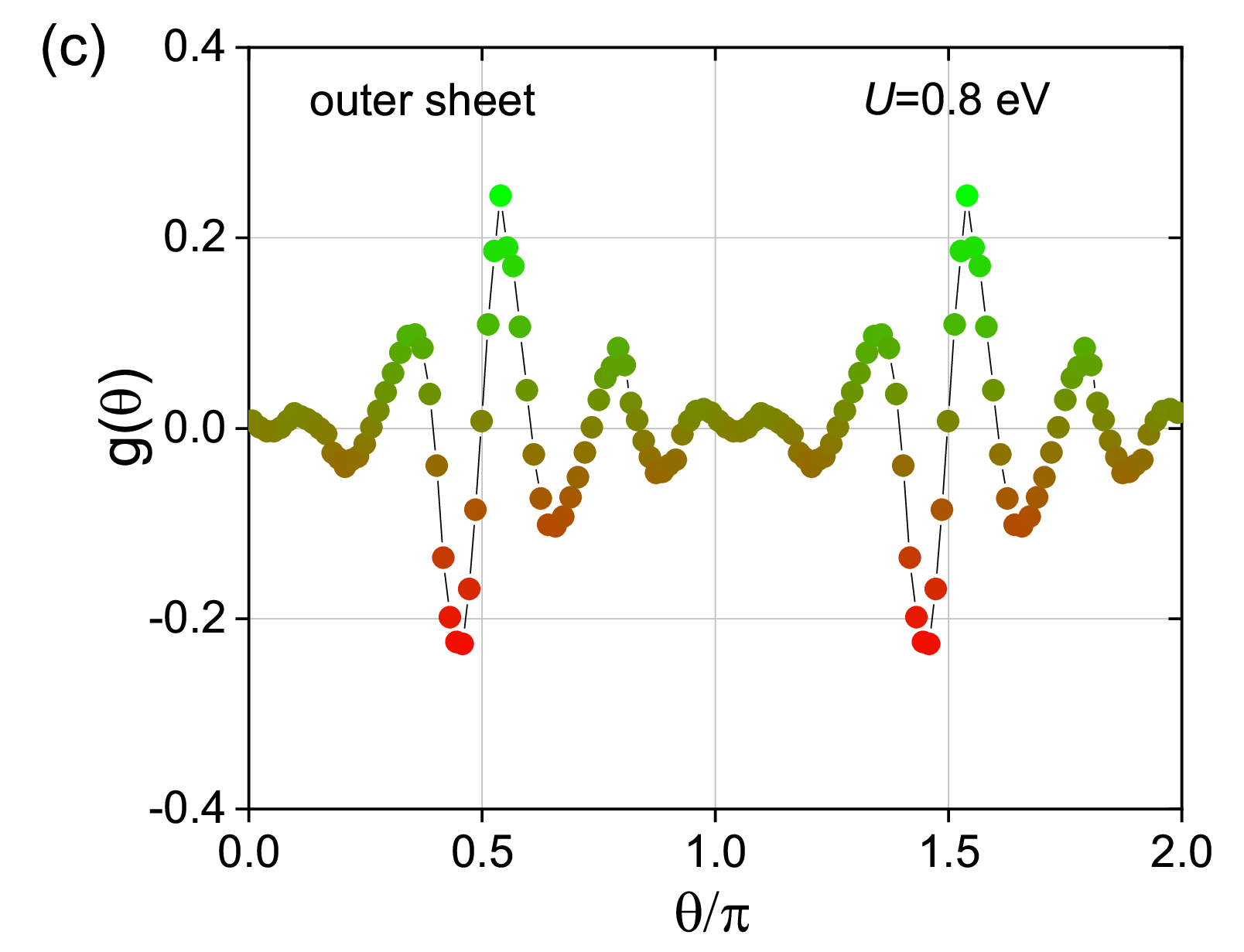}
\includegraphics[angle=0, width=0.49 \columnwidth]{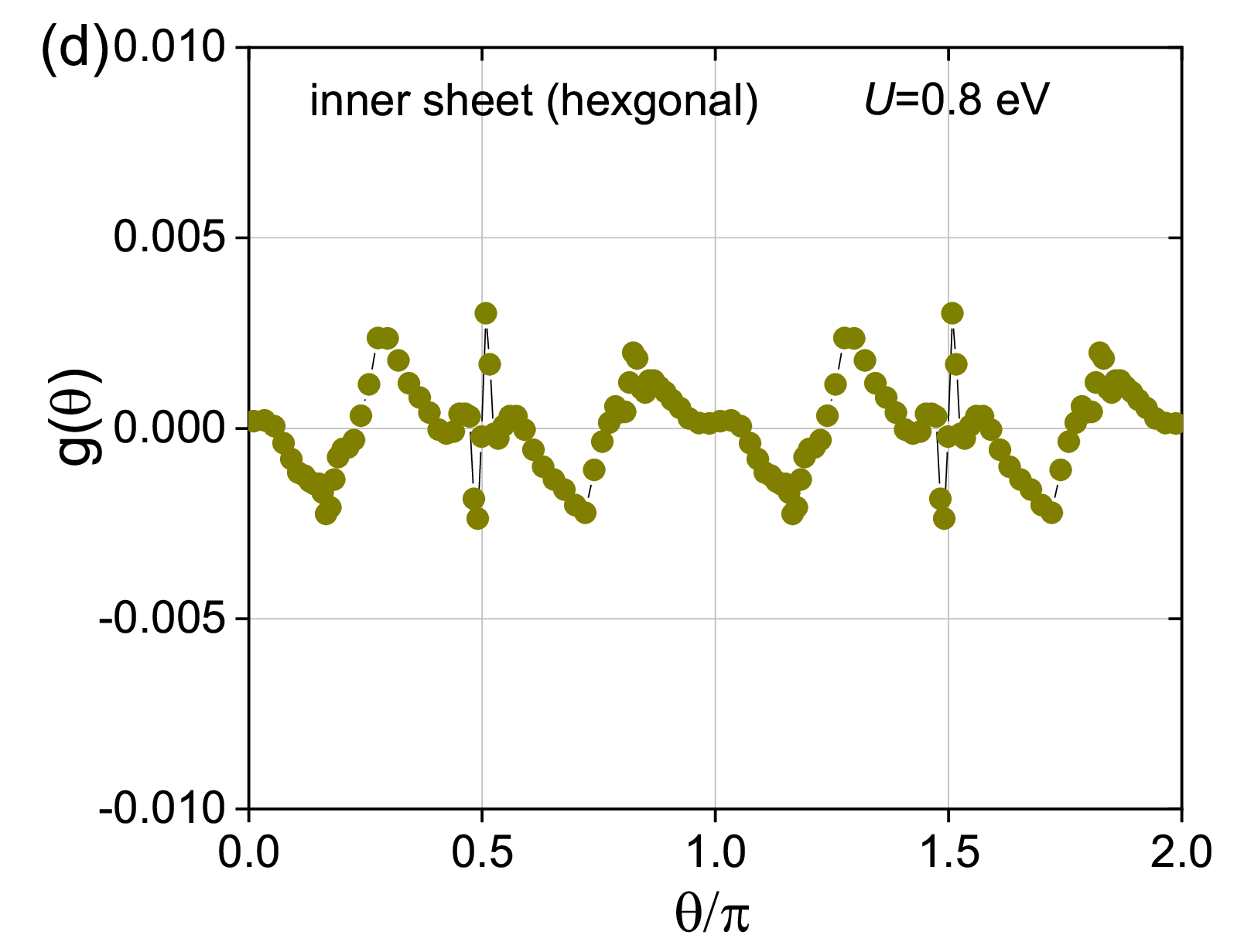}
	\caption{The dominant singlet pairing gap function $g(k)$ versus k around Fermi surface at $U=0.5$ eV (a),(b) and 0.8 eV (c),(d), on the outer and inner Fermi surfaces, respectively.} 	   
	\label{Fig.gap_theta}
\end{figure}

\noindent{\it Superconductive Pairing Strengths}:
By solving the linearized gap eqaution (\ref{eq:linearized gap equation}), we obtain the largest singlet and triplet pairing strengths as a function of the Coulomb interaction $U$. The results shown in Fig. \ref{Fig.pairing} indicate the singlet pairing dominates in the parameter range of $U>0.4$ eV. This suggests that the Cooper pairs in CsV$_{3}$Sb$_{5}$ are singlet pairing. It is consistent with the decrease of Knight shift of Sb with decline of temperature\cite{ChaoMu-77402}. Meanwhile, we find that there exists the distinct difference in the SC gap functions on the two Fermi surfaces, the SC pairing gap function on inner Fermi surface is almost two order smaller in magnitude than that on outer Fermi surface.
\\

The largest singlet pairing gap functions g(k) at $U=0.5$ eV and 0.8 eV are plotted in Fig. \ref{Fig.gap function}. These gap functions show sign-change at the high symmetry k points on the Fermi surface. 
Further analysis in Fig. \ref{Fig.gap_theta} shows when $U=0.5$ eV,  the gap function exhibiting the antisymmetry with respect to the $x$ axis and the $y$ axis, {\it i.e.}, d$_{xy}$-wave-like symmetry, on the outer Fermi sheet and weak g-wave symmetry on the inner Fermi sheet. The gap function on the inner Fermi sheet is about smaller than that on the outer Fermi sheet by two orders in magnitude.  When $U=0.8$ eV, the 
gap functions slightly change, while still are d$_{xy}$-wave-like.
These suggest that the d$_{xy}$-wave-like paring is dominant in CsV$_{3}$Sb$_{5}$. The d$_{xy}$-wave-like pairing 
is in consistent with the results of finite linear term of thermal conductivity at ultra low temperature\cite{zhao2021nodal} and the V-shaped gap\cite{nature-222-599}, which suggest the nodal superconductivity in CsV$_{3}$Sb$_{5}$. However, the study of impurity effects\cite{PhysRevLett-127-187004} on superconductivity by STM and the appearance of Hebel-Slichter peak\cite{ChaoMu-77402} in $1/T_{1}T$ just below T$_{C}$ proposes S-wave pairing in CsV$_{3}$Sb$_{5}$. To uncover the structure of the superconducting gap in CsV$_{3}$Sb$_{5}$, further experiments on superconducting state, such as angular-resolved photoemission spectroscopy (ARPES), Bogoliubov quasiparticle interference, and phase-sensitive tests, are crucial and highly expected.
\\

%CCCCCCCCCCCCCCCCCCCCCCCCCCCCCCCCCCCCCCCCCCCCCCCCCCCCCCCCCCCCCCCCCCCCCCCCCCCCCCCCCCCC
																							
%\section{Discussions and Conclusion}
%\label{summary}

In fitting the present effective model (1), we elaborately consider the major bands contributed from the 3d orbits of V ions and ignore the bands contributed from Sb 5p orbits. The latter consists of the central Fermi surface around the $\Gamma$ point which arises from Sb 5p orbits and a few of Fermi surface fragments arising from the hybridization between dominant Sb 5p orbits and partial V 3d orbit.
One notices an experimental fact that upon applying pressure, the central Fermi surface and Sb-related 5$p$ bands of CsV$_3$Sb$_5$ almost do not vary with the disappearance of the charge-density-wave order \cite{PhysRevX-11-041010}, while the superconducting properties change considerably with increasing pressure, suggesting that the present fitted effective minimal model is responsible for the unconventional superconductivity in CsV$_3$Sb$_5$. 
\\

Contrast to the present singlet d$_{xy}$-like superconductive pairing symmetry in our effective minimal model for CsV$_3$Sb$_5$, very recently
Wu {\it et al.} fitted another effective 6-orbital model \cite{PhysRevLett-127-177001} based on the electronic structures of KV$_3$Sb$_5$, they found that within the RPA their model favors the triplet f-wave pairing symmetry.
%
% An earlier study by Kang {\it et al.} \cite{Kang-2011} about the superconductivity of a single-band Hubbard model on a Kagome lattice 
% showed that the pairing nature is spin singlet and d-wave-like symmetry, similar to the present results.
%
This indicates that the superconductive pairing symmetry is sensitive to the details of the band structures in AV$_3$Sb$_5$. Our present singlet pairing results agree with recent experiments, suggesting the plausible of the present effective minimal model for the superconductivity in AV$_3$Sb$_5$.
\\

In summary, we have obtained an effective low-energy six-band model for CsV$_3$Sb$_5$ by fitting the the first-principles electronic structures. Within the random phase approximation and on the basis of the effective minimal model, we have found that the superconducting pairing strength increases with the lift of Coulomb correlation and decline of temperature,
and superconductive pairing symmetry is singlet, most probably with the d$_{xy}$-wave-like symmetry. These
results highlight the essence of the unconventional superconductivity in Kagome compounds AV$_3$Sb$_5$.
Also, one may recall that the full effective model should include the central Fermi surface and a few of Fermi surface fragments ignored in the present study,  the role of these dominant Sb 5p bands on the nature of the superconductivity in Kagome compounds AV$_3$Sb$_5$ is worthy of further investigations. 

%ccccccccccccccccccccccccccccccccccccccccccccccccccccccccccccccccccccccccccccccccccccccccccccccccccccc

\begin{acknowledgements}
%{\it Acknowledgements}：
The authors thank the supports from the NSFC of China under Grant nos. 11774350 and 51727806, 
Program of Chinese Academy of Sciences, Science Challenge Project No. TZ2016001. Numerical 
calculations were partly performed at the Center for Computational Science of CASHIPS,
the ScGrid of the Supercomputing Center, the Computer Network Information Center of CAS, 
the CSRC computing facility, and Hefei Advanced Computing Center.
\end{acknowledgements}
\noindent{\it Note added.} After we finished the manuscript, we realized that Lin {\it et al.} suggested that the Pomerranchuk  fluctuations may lead to the  superconductivity in the Kagome metals AV$_3$Sb$_5$ \cite{lin-2021-kagome}. Whether the van Hove singularity related Pomeranchuk fluctuations could survive in the present intermediate correlated compounds deserves further study.

%\nocite{*}
%\bibliography{apssamp}

\end{document}